\documentclass[11pt]{article}

\usepackage{amsmath}
\usepackage{amsthm}
\newtheorem{theorem}{Theorem}[section]

\newtheorem{lemma}[theorem]{Lemma}
\newtheorem{define}[theorem]{Definition}

\usepackage{epsfig}
\usepackage{graphicx}

\newcommand {\E}{\mathbf E}
\newtheorem{claim}[theorem]{Claim}

\newcommand{\EE}{\epsilon}
\newtheorem{ex}{Example}[section]

\newcommand{\eat}[1]{}
\renewcommand{\paragraph}[1]{\medskip \noindent {\bf{#1}}}

\advance \topmargin by -\headheight
\advance \topmargin by -\headsep
\evensidemargin \oddsidemargin
\begin{document}
\title{Adaptive Uncertainty Resolution in Bayesian Combinatorial Optimization Problems\thanks{Parts  of this paper appeared in the Proceedings of the $18^{th}$ ACM-SIAM Symposium on Discrete Algorithms, 2007~\cite{GuhaM07}. \newline
Author's address: Sudipto Guha, Department of Computer and Information Sciences,University of Pennsylvania. Supported in part by an Alfred P. Sloan Research Fellowship, by NSF awards CCF-0644119, and CNS-0721541. Email:  {\tt sudipto@cis.upenn.edu}. \newline
Kamesh Munagala, Department of  Computer  Science, Duke  University, Durham NC 27708. Research supported by an Alfred P. Sloan Research Fellowship, and by NSF via a CAREER award and grant CNS 0540347. Email: {\tt kamesh@cs.duke.edu}.}}
 \author{Sudipto Guha\\University of Pennsylvania\and
Kamesh Munagala\\Duke University}
 
 \maketitle

\begin{abstract}
  In several applications such as databases, planning, and sensor
  networks, parameters such as selectivity, load, or sensed values are
  known only with some associated uncertainty. The performance of such a
  system (as captured by some objective function over the parameters) is
  significantly improved if some of these parameters can be probed or
  observed. In a resource constrained situation, deciding which parameters
  to observe in order to optimize system performance, itself becomes an
  interesting and important optimization problem. This general problem is
  the focus of this paper. 

  One of the most important considerations in this framework is whether
  adaptivity is required for the observations. Adaptive observations
  introduce blocking or sequential operations in the system whereas
  non-adaptive observations can be performed in parallel. One of the
  important questions in this regard is to characterize the benefit of
  adaptivity for probes and observation.

  We present general techniques for designing constant factor
  approximations to the optimal observation schemes for several widely
  used scheduling and metric objective functions. We show a unifying
  technique that relates this optimization problem to the {\em outlier}
  version of the corresponding deterministic optimization. By making this
  connection, our technique shows constant factor upper bounds for the
  benefit of adaptivity of the observation schemes. We show that while
  probing yields significant improvement in the objective function, being
  adaptive about the probing is not beneficial beyond constant factors.
\end{abstract}

\section{Introduction}
\label{sec:intro}
Consider a measurement scenario such as sensor networks or spectrograms,
where we have errors in estimation, but the distribution of the error is
understood through models of the measuring instrument, historical data,
regression, Kalman filters, etc.  For example, suppose we have a set of
data-points and are trying to construct a classification of them using the
minimum spanning tree (MST).  Suppose that we cannot resolve all the
errors but we can repeat and refine a few of the measurements, i.e,
``probe'' a few points to make detailed measurements. A natural question
is which measurements do we refine, say if we wanted to refine only $k$ of
them and wanted to optimize the expected cost of the MST?  In many
situations, we can assume that the errors made by different instruments
are independent of each other -- even then the problem is {\sc
  NP--Hard}~\cite{GoelGM06} for computing very simple functions such as
the {\em minimum} of the readings. In this paper we focus on general
combinatorial optimization problems.

The above is a typical example of a model driven optimization
problem~\cite{GoelGM06}.  This has gained significant currency in a
variety of other research areas such as database query optimization and
route selection in networks.  In a database query optimization setting
suppose the optimizer is presented with a set of sufficiently complicated
and unrelated (independent) queries. The query-optimizer can estimate the
resources needed by the queries from historical information, cached
statistics, sampling of various sub-queries, or by performing inexpensive
filters~\cite{Shiv05,ChuHG02,ChuHS99}. Subsequent to this estimation, the
query optimizer schedules the tasks to optimize the throughput or the
average completion times.  However, this estimation process itself
consumes resources such as time, network bandwidth, and space, and
therefore the decision to choose the parameters to refine in estimate
becomes a key optimization problem.  A similar problem arises in
networking, where the current state of multiple routes and servers can be
probed and observed before deciding which route/server to use for a
specific connection~\cite{AkellaMSSS03,GummadiMGLW04,GuhaMS06}.  The above
examples can also be extended to planning over a network, where we may
wish to visit a sequence of nodes, about which we have imprecise
information and can only refine a small number of them, to minimize total
distance traveled -- such problems arise in query processing in sensor
networks~\cite{Deshpande}.

\renewcommand{\v}{\mathbf{v}}
\renewcommand{\O}{\mathcal{O}}
\newcommand{\ol}{\bar}

\paragraph{Problem Statement.} The above problems of measurement refinement or optimizing independent query schedules can be formulated abstractly as follows. Let $X_1, X_2, \ldots, X_n$ be non-negative independent random variables, whose distributions are given as inputs (the distribution can be specified by samples and does not affect any result in this paper). Let $[n]$ denote $\{1,2,\ldots,n\}$. For subset $S \subseteq [n]$, let $\v_{S}$ denote the realization of $\{X_i, i \in S\}$. We are given a minimization problem $h(o; \v)$, where $\v \in \Re_+^{n}$ denotes the inputs, and $o \in \O$ denotes the output.   

As an example, in weighted completion time scheduling on a single machine,  $1||\sum_j w_j C_j$, job $i$ has size that follows distribution $X_i$, and has weight $w_i$.  The vector $\v$ denotes a realization of job sizes, the space $\O$ is the set of all orderings of jobs, and the function $h(o; \v) = \sum_i w_i \sum_{j \prec_o i} \v_j$, where $j \prec_o i$ means job $j$ appears before $i$ in the ordering $o$. Note that if the goal is to solve $\min_{o \in \O} \E_{\v}[h(o;\v)]$, the optimal $o$ simply orders the jobs by Smith's rule in decreasing order of $\frac{w_i}{\E[X_i]}$. If the goal is instead to solve $\E_{\v}[\min_{o \in \O} h(o; \v)]$, for each realization $\v$, the ordering would be in decreasing $\frac{w_i}{\v_i}$. Our goal will not be to solve either of these extreme versions.

Our goal will be to solve the probing version of these problems.  Here, each variable $i \in [n]$ has an observation cost $c_i$. The exact value of the underlying random variable $X_i$ can be found by spending this cost. There is a budget $C (\ge \max_i c_i)$ on the total cost of observation. The probing policy chooses a subset of variables of total cost at most $C$ to probe and observe. {\em Subsequent} to this, the optimization policy chooses a solution based on the outcome of the probes.   We note that the solution is fixed {\em after} the outcomes of the probes, but {\em before} the realization of the unprobed variables. Therefore, given the set of probed variables and their outcomes, the solution is the same for {\em all} realizations of the unprobed variables. The final goal is to
devise a policy (or strategy) for adaptively choosing that $S \subseteq [n]$ of variables so that $\E_{\v_{S}}\left[\min_{o \in \O} \E_{\v_{\ol{S}}}\left[h(o; \v_{S}, \v_{\ol{S}})\right]\right]$ is minimized. The expectation is jointly over outcomes of probes and the distributions of unprobed variables. Here, $\v_S$ denotes the realization of the random variables $\{X_i, i \in S\}$, and $\ol{S}$ denotes $[n]\setminus S$.  

In the above completion time example,  let $\E[X_i] = \mu_i$. If the probing policy chooses a subset $S$ of jobs to probe, after the probing, the optimal ordering is Smith's Rule using the exact sizes $\v_i$ for the probed jobs and the expected sizes $\mu_i$ of the unprobed jobs. For a given realization $\v_{S}$ of the variables $\{X_i, i \in S\}$, the expected completion time (where the expectation is over the unprobed job sizes) is:
\begin{equation*}
\begin{split}
\min_{o \in \O} \E_{\v_{\ol{S}}}\left[h(o; \v_{S}, \v_{\ol{S}})\right]  & = \label{eq}\sum_{i, j \in S, j \ge i} \min(w_i \v_j, w_j \v_i)  \\  &+ \sum_{i,j \notin S, j \ge i} \min(w_i \mu_j, w_j \mu_i) + \sum_{i\in S, j \notin S} \min(w_i \mu_j, w_j \v_i) 
\end{split}
\end{equation*}
The value of the overall solution given $S$ is probed is therefore:
\begin{equation}
\E_{\v_S}\left[\min_{o \in \O} \E_{\v_{\ol{S}}}\left[h(o; \v_{S}, \v_{\ol{S}})\right]\right]
\label{eq1}
\end{equation}
 The objective of the probing strategy is to adaptively choose a subset $S$ to probe, so that this value is minimized, subject to $\sum_{i \in S} c_i \le C$. Note that such a strategy is a decision tree with the next decision to probe depending on the outcomes so far. At a leaf $l$ of the decision tree, a subset $S_l$ has been probed and realized; the job ordering at the leaf would have weighted completion time given by Equation (\ref{eq}) with $S = S_l$. The expected value of this quantity over the decision tree is the value of the probing strategy.  The goal is to design a polynomial time algorithm for finding the optimal probing strategy.

\paragraph{Conceptual Issues.} Several conceptual and systems issues arise immediately.
\begin{itemize}
\item The most important consideration for a system is quantifying the
  benefit of being adaptive in the probing policy.  In the above
  formulation, the probes could possibly all be done in parallel upfront,
  {\em i.e.}, non-adaptively, and the results of the probes are ignored in
  deciding which other variables to probe. In contrast, adaptive
  observations are sequential, based on the outcomes of the previous
  probes. It is clear that adaptive strategies yield as good or better
  solutions than the non-adaptive counterpart. The first issue with an
  adaptive strategy is the complexity of expressing the solution; it is
  {\em a priori} not clear that the decision tree that encodes the optimum
  probing policy (or strategy) is polynomially bounded, so that it is not
  clear if the problem of finding the optimal adaptive probing policy is
  even in {\sc NP}.

\item The next interesting issue, which is a consequence of the adaptivity, is the budget. Note that while probing, if a suitable value is found already, then the probing can be halted early, yielding significant cost savings. Thus it makes sense to assume that the budget is in expectation as well. This sets up two classes of problems, namely with {\em hard budgets}, where the probing budget is $C$ is on all decision paths,  and with {\em soft budgets}, {\em i.e.}, the budget $C$ is in expectation over the decision paths, so that on some paths, the cost of the variables probed could exceed $C$ if this is compensated by halting early on other decision paths. Note that this distinction does  not arise for non-adaptive policies, where the probes are done upfront in parallel. 
\end{itemize}

\paragraph{Results.}
 In this paper, we design poly-time algorithms for finding approximately optimal {\em non-adaptive} probing strategies for the large class of scheduling and metric objective functions that have been considered in the known applications and literature to date, namely, weighted completion time scheduling, minimum makespan scheduling, metric clustering, and Steiner trees.  Our results for all  these seemingly disparate problems use a common underlying technique that also shows that these strategies have a bounded {\em adaptivity gap}, i.e., there exists a non-adaptive probing strategy which is only a constant factor worse compared to the best adaptive probing strategy with soft budgets. Therefore for the problems we study, the non-adaptive and adaptive models with hard and soft budgets are all related to each other by constant factors. As a consequence, one of the main points we  establish is that although probing helps significantly\footnote{We show in Section~\ref{avgcompletion} that probing does indeed improve the objective of $1||\sum_j w_j C_j$ by a polynomial amount.}, adaptive probing is no better than non-adaptive probing by more than a constant factor in many problems of interest.  

\paragraph{Related Work.}  The model-driven optimization framework can be defined for other objective functions $h$. This class of problems was first defined by the authors of~\cite{GoelGM06}, where the non-adaptive {\sc minimum-element} problem, with $h = \min_i X_i$ is considered. Though the current paper presents the best results known for scheduling and metric problems, other work~\cite{GoelGM06,GuhaM07,GuhaM07b,GuhaMS06} has considered different objective functions, most notably problems involving subset selection. In~\cite{GuhaM07,GuhaM07b}, we considered the case where $f$ is a single constraint packing problem such as knapsack with random profits which are observable, and present a $8$ approximation based on rounding the solution of a natural linear program. As in this paper, the approximation ratio holds even when the optimal solution is allowed adaptive ({\em i.e.}, can be based on the results of previous observations). It further holds even when the hidden quantity is a distribution (instead of a single value) and a prior on this distribution is specified as input.  In~\cite{GuhaMS06} the authors consider the Lagrangean version of $h = \max_i X_i$, where the observations are adaptive and the goal is to maximize the expected difference between the maximum value and the observation cost. Note that there is no budget on this cost, instead it is part of the objective function.  We note that the techniques needed for subset selection problems are very different from those needed for the scheduling and metric problems considered in this paper, where all variables whether probed or unprobed eventually are part of the solution.

The notion of refining uncertainty has been considered in an {\em
  adversarial setting} by several
researchers~\cite{OlstonThesis,FederMPOW00,KhannaT01,CharikarFGKRS00}.
Here, the only prior information about an input is the lower and upper
bounds on its value. The goal is to minimize the observations needed
to estimate some function over these inputs {\em exactly}, and often
strong lower bounds arise.  For correlated random variables, the
problem of minimizing residual information is considered
in~\cite{krause}.

\medskip
The adaptivity gap in the absence of probes has been considered in the literature earlier, notably in~\cite{DeanGV04,DeanGV05,MohringSU,SkutellaU} for Knapsack and scheduling problems.  However, the model driven optimization problem is considerably different from the settings in those papers.  In \cite{DeanGV04,DeanGV05,MohringSU,SkutellaU} the optimum is allowed to decide on the next item to schedule based on the past, but once the next item is decided this is an irrevocable commitment.  In contrast, in our problem, after probing we (as well as the optimum adversary) may choose {\em not to} include an item in the Knapsack, {\em arbitrarily (re)order} the schedule, come up widely different clustering depending on the outcome of the probed values.  Hence,  we need completely different arguments and observation schemes in the absence of the irrevocable commitment.  
 
The classic stochastic optimization (non-adaptive, non-probing) versions of these scheduling problems were considered in~\cite{KleinbergRT97,GoelI99}.    We also note that stochastic optimization problems were considered in~\cite{ImmorlicaKMM04,GuptaPRS04,GuptaRS04,ShmoysS04,ShmoysS05,root,wednesday}; these problems appear to be unrelated to adaptivity gaps.

\section{General Framework}
The types of objectives we are interested in this paper are scheduling and
metric problems. For these problems, the key feature is that the final
solution has to be constructed over {\em all} the variables whether probed
or unprobed.  Our main contribution is to identify a key {\bf recombinant}
property  and an uniform solution recipe that shows
constant factor adaptivity gap for all problems which satisfy the property. We present this general technique below, and adapt it to specific problems in the subsequent sections.

\subsection{Problems Considered}
Before presenting the general algorithmic framework and analysis using the recombinant property, we summarize our problems and results. 

\medskip\noindent{\bf Scheduling Problems.} For scheduling we consider the average completion time ($1||\sum_j w_j C_j$) and minimum makespan on identical machines; in both cases, the job sizes are random variables. We focus on the weighted completion time which illustrates the main issues (Section~\ref{avgcompletion}). A surprising feature of this algorithm is that {\em only the expected values} of job sizes are used in constructing the approximately optimal probing scheme. The makespan problem is discussed in Section~\ref{app:makespan}. For both these problems we design non-adaptive algorithm which shows that the adaptivity gap is a constant. 

\medskip\noindent{\bf Metric Problems.} We next consider several optimization problems in metric spaces. We assume that the input nodes are discrete distributions of polynomial specification over points of the metric space. For these problem a small but  important added twist is needed -- we need to reformulate the problem on a different but related metric.  We consider the $k$-Median and Steiner Tree (or TSP) problem. We discuss the $k$-Median problem and the general setup for these problems
in Section~\ref{kmedian}. The MST problem is discussed in Section~\ref{app:mst}. For both these problems we design non-adaptive algorithm which shows that the adaptivity gap is a constant.

For the $k$-median problem an interesting issue comes to fore, which does not appear for deterministic input. It is well known that over any space, there exists a $k$-median solution which uses the input points and is at most twice the optimum. For distributional input we show that the adaptivity gap for obtaining exactly $k$ medians is polynomially large, short of a polynomial blowup in the probing cost.  However, if we restrict all solutions to use fixed points in the metric space as medians (as opposed to declaring an input node which could possibly be a distribution over points as a median) then the gap disappears! This exposes an interesting contrast in the problem based on which points are allowed to be medians. We expect these definitional issues to be of independent interest as more problems with distributions as input are investigated.

\subsection{The Non-adaptive Algorithm}  For $S \subseteq [n]$, let $\O_{S}$ denote the space of solutions constructed on the subset $S$ of input variables. Let $\O = \O_{[n]}$. For instance, in weighted completion time scheduling, $\O_S$ denotes the space of orderings of the jobs in subset $S$. Given a realization $\v_A$ of variables $\{X_i, i \in A\}$, let $\min_{o \in \O_A} h_A(o,\v_A)$ denote the optimal objective using only these realized variables.

\newcommand{\G}{\mathcal{G}}

\begin{define}
The {\em outlier} problem $\G^*(C)$ is: Choose a subset $S$ of variables so that $\sum_{i \in S} c_i \le C$ as outliers ({\em i.e.}, to ignore), so that $\min_{o \in \O_{\ol{S}}}\E_{\v_{\ol{S}}}\left[h_{\ol{S}}(o; \v_{\ol{S}})\right]$ is minimized. Here, $\ol{S} = [n] \setminus S$. 
\end{define}

As an example, in weighted completion time scheduling, $\G^*(C)$ would correspond to choosing a subset $S$ of jobs with total probing cost at most $C$ as outliers ({\em i.e.}, to ignore) so that for the remaining jobs $\ol{S}$, the expected completion time (with no probing), $\sum_{i,j \in \ol{S}, j \ge i} w_j \mu_i$ is minimized. Here, $\mu_i = \E[X_i]$. 

\medskip
Our non-adaptive strategy has the following simple structure: 
\begin{enumerate}
\item Solve the {\em outlier} problem $\G^*(C)$. Suppose this outputs a subset $S \subseteq [n]$ of  as outliers. 
\item The non-adaptive algorithm probes $S$, and for the realization $\v_{S}$ of these variables, subsequently solves $\min_{o \in \O} \E_{\v_{\ol{S}}}\left[h(o; \v_S, \v_{\ol{S}})\right]$ on the observed and unobserved variables.
\end{enumerate}

As an example, in weighted completion time scheduling, suppose the outlier problem returns a subset $S$ of jobs as outliers. The probing strategy observes the sizes of these jobs. It then sorts the jobs in decreasing order of $\frac{w_i}{l_i}$ and schedules them, where $l_i = \mu_i$ if $i \in \ol{S}$, and $l_i = \v_i$ if $i \in S$. 

\subsection{Analysis}  
The guarantees we obtain will depend on the approximation ratio achievable for $\G^*(C)$ and for solving $\min_{o \in \O} \E_{\v_{\ol{S}}}\left[h(o; \v_S, \v_{\ol{S}})\right]$.  We will require two properties about the problems we consider. The first property is {\em downward closure}.
\begin{equation*}
\forall  o \in \O_{[n]}, A \subseteq [n],  \qquad \exists o_A \in \O_A \ \   s.t. \qquad \forall \v_{[n]} \qquad h_A(o_A; \v_A) \le h(o; \v_{[n]}) \tag{P1}
\end{equation*}

The above property trivially holds for all the problems we consider: weighted completion time scheduling and minimum makespan scheduling where $\v$ denotes job sizes; and geometric $k$-medians and Steiner trees, where $\v$ denotes locations of points in a metric space. 

 Using (P1), we will show a lower bound for the adaptive optimal solution in terms of the outlier problem. In particular, we relate the quantity $\G^*(C)$ to the values $OPT_s$ and $OPT_h$ of the optimal adaptive solutions with soft and hard probing budgets $C$ respectively. Recall that in the hard budget version, the probe cost has to be at most $C$ on all adaptive decision paths, whereas in the soft budget version, it is only required that the expected probe cost be at most $C$. Clearly, $OPT_h \ge OPT_s$ since the soft budget version relaxes the hard budget constraint. 

\begin{lemma} 
\label{lem:approx}
 $\G^*(C) \le OPT_h$, and $ \G^*((1+\beta)C) \le \left(1+\frac{1}{\beta} \right) OPT_s$ for all $\beta > 0$.
\end{lemma}
\begin{proof}
First consider the optimal decision tree with hard budgets. At each leaf $l$ of this tree, the probing cost is at most $C$. Let $S$ denote the subset of probed variables corresponding to this leaf node, and $\v_S$ denote the values observed for these variables.  Let $opt_l$ denote the objective function at this leaf node. By property (P1), we have:
$$ opt_l  =  \min_{o \in \O} \E_{\v_{\ol{S}}}\left[h(o; \v_S, \v_{\ol{S}})\right]  \ge   \min_{o_{\ol{S}} \in \O_{\ol{S}}} \E_{\v_{\ol{S}}}\left[ h(o_{\ol{S}};  \v_{\ol{S}})\right]   \ge  \G(C^*) $$
Therefore, $opt_l \ge \G^*(C)$. Since this is true for all leaves $l$, we have $OPT_h \ge \G^*(C)$.

Next consider the optimal decision tree with soft budget $C$. Consider all leaves $l$ with probing cost at most $(1+\beta) C$. By Markov's inequality, the probability of such a leaf is at least $\frac{\beta}{1+\beta}$. For each of these leaves, $opt_l \ge \G^*((1+\beta) C)$. Therefore using (P1), we have:
$$OPT_s \ge \frac{\beta}{1+\beta} \G^*((1+\beta) C) \qquad \Rightarrow \qquad \G^*((1+\beta) C) \le \left(1+\frac{1}{\beta} \right)  \cdot OPT_s$$
\end{proof}

\paragraph{Recombinant Property.} The next and more non-trivial property relates to combining the solutions for the probed and unprobed parts. This property crucially requires that the random variables $X_i$ are independent, and holds for all product distributions. In the inequality below, the expectations are over $\v_{\ol{A}}$. 

\begin{equation*}
\begin{split}
\forall A \subseteq [n], \v_A, & \min_{o \in \O} \E\left[h(o; \v_A, \v_{\ol{A}})\right]  \\ \le & \rho\left(\min_{o \in \O_A} h_A(o; \v_A)
+ \min_{o \in \O_{\ol{A}}} \E[h_{\ol{A}}(o; \v_{\ol{A}})] + \E[\min_{o \in \O} h(o; \v_A, \v_{\ol{A}})] \right) \end{split}
\tag{P2} 
\end{equation*}

At a high level, the recombinant property holds if for an arbitrary partitioning of the input variables, each of the parts have an induced solution with the property that the objective function values sum to no more than the respective quantities in the original problem; further, the two solutions can be combined without a significant increase in value of the resulting solution. Since encoding the value of the probed part is unwieldy, this partitioning is necessary to construct a solution that encodes just the unprobed part. The recombination shows a small adaptivity gap -- the tricky part is to ensure that the interaction of the probed and unprobed parts can be bounded. This technique also yields non-adaptive probing strategies in addition to the adaptivity gap proof.

For all the problems we consider, we will show that $\rho$ is a constant. Using this, we can now derive the approximation ratio for the non-adaptive strategy. Suppose Step (1), the outlier problem, has approximation ratio $\gamma_1$, and Step (2), solving  $\min_{o \in \O} \E_{\v_{\ol{S}}}\left[h(o; \v_S, \v_{\ol{S}})\right]$ given $\v_S$ has approximation ratio $\gamma_2$, then we have the following result:

\begin{theorem}
The non-adaptive algorithm finds a solution of value at most $\rho \gamma_1 (2+\gamma_2) OPT_h$. It further finds a solution of value $\rho \gamma_1 (2+\gamma_2) \left(1+\frac{1}{\beta} \right) OPT_s$ using outlier cost $(1+\beta) C$ for any $\beta > 0$.
\end{theorem}
\begin{proof}
Given the computed set of outliers $S$ and the final solution $\nu_{\v_S}$ for the hard budget case, we have the following sequence of inequalities:
\begin{eqnarray*}
&&\E_{\v_S} \left[ \E_{\v_{\ol{S}}}\left[h(\nu_{\v_S}; \v_S, \v_{\ol{S}})\right]\right] \\
&& \le  \gamma_2 \E_{\v_S} \left[\min_{o \in \O} \E_{\v_{\ol{S}}}\left[h(o; \v_S, \v_{\ol{S}})\right]\right]\\
&& \le   \gamma_2 \rho \E_{\v_S} \left[  \min_{o \in \O_S} h_A(o; \v_S) + \min_{o \in \O_{\ol{S}}} \E[h_{\ol{S}}(o; \v_{\ol{S}})] + \E[\min_{o \in \O} h(o; \v_S, \v_{\ol{S}})] \right] \\
&& \le  2 \gamma_2 \rho \E_{\v} \left[\min_{o \in \O} h(o; \v)\right] + \gamma_2 \rho \E_{\v_S} \left[ \min_{o \in \O_{\ol{S}}} \E[h_{\ol{S}}(o; \v_{\ol{S}})] \right]\\
&& \le   2 \gamma_2 \rho \E_{\v} \left[\min_{o \in \O} h(o; \v)\right] + \gamma_2 \rho \gamma_1 \G^*(C) \\
&& \le  \gamma_2 \rho (\gamma_1 + 2) OPT_h
\end{eqnarray*}
The second inequality follows from Property (P2), the third from Property (P1), and the final inequality from Lemma~\ref{lem:approx}. The proof for the soft budget case is identical.
\end{proof}

\begin{define}
We will denote $\eta = \gamma_2 \rho (\gamma_1 + 2)$ as the {\em approximation ratio} of the problem.
\end{define}

The above guarantees imply bicriteria approximations (where  the probing cost also increases by a constant factor) for the soft budget versions of all problems, and for the hard budget versions of the metric problems we consider. The increase in probing cost is unavoidable using our techniques, since the  best known algorithms for the outlier versions of the metric problems we consider have similar gaps. An interesting open question  is to show a complexity result  that the increase in budget is unavoidable. 

\section{Scheduling I: Weighted Completion Time on Single Machine}
\label{avgcompletion}
We first consider the weighted completion time problem ($1||\sum w_j C_j$) of scheduling jobs on a single processor to minimize the sum of the weighted completion times.  All jobs are released at time $t = 0$ and there are no deadlines or precedence constraints. 

In the probing model the sizes (or processing times)  of jobs $J_1,\ldots, J_n$  are distributed according to independent random variables $X_1,X_2,\ldots,X_n$ respectively. The weight of job $J_i$ is $w_i$, which is {\em not} a random variable. Let $\E[X_i] = \mu_i$. Each variable $X_i$ corresponding to the size of job $J_i$ has probing cost $c_i$; probing yields its exact value.  Let $C$ denote the (soft) budget on probing cost.

The solution is a strategy for adaptively probing a subset the jobs so that the expected weighted completion time of scheduling {\em all} the jobs after the outcome of the probes is known, is minimized. This expectation is over the outcome of the probes, and over the distribution of the processing times of the unprobed jobs.   We note that the scheduling policy fixes the ordering $o \in \O$ of all jobs after the results of the probes (say set $S$) are known, but before the sizes of the unprobed jobs are revealed. Therefore, the optimal scheduling policy whose value is $\min_{o \in \O} \E_{\v_{\ol{S}}}[h(o; \v_S, \v_{\ol{S}})]$ will simply order the jobs in decreasing order of weight to the ratio of (expected) processing time ({\em Smith's Rule}) --  this processing time is exactly known for probed jobs $S$, and is the expected processing time  for unprobed jobs $\ol{S}$\footnote{This is verified by estimating the benefit/loss of an exchange of consecutive jobs on each scenario and then aggregating over the scenarios, see also \cite{pinedo}.}. 

\medskip
\noindent {\bf Benefit of Probing.} To gain intuition, we present a simple example where probing helps by a factor of $\Omega(n)$. There are $n$ jobs with unit weights and unit probing costs, and sizes which are $0$ w..p. $1-1/n$, and $1$ with the remaining probability. If no jobs are probed, each job has expected size $1/n$, so that the expected completion time of any ordering is $\sum_{i=1}^n i/n = \frac{n+1}{2}$. If all jobs are probed, the jobs which are of size $1$ can be placed after jobs of size $0$. If there are $k$ jobs of size $1$, the completion time is $\frac{k(k+1)}{2}$. The variable $k$ follows Binomial$(n,1/n)$, with mean $1$ and variance at most $1$. This implies $\E[k^2] \le 2$, so that $\E\left[\frac{k(k+1)}{2}\right] \le \frac{3}{2}$. Therefore, probing yields a $\Omega(n)$ benefit. 

\paragraph{Main Result.} Recall that $\rho$ is the approximation for Property (P2), $\gamma_1$ is the approximation ratio for the outlier problem $\G(C^*)$, and $\gamma_2$ is the approximation ratio for solving $\min_{o \in \O} \E_{\v_{\ol{S}}}\left[h(o; \v_S, \v_{\ol{S}})\right]$. We will show the following theorem.
 
 \begin{theorem}
 \label{thm1}
 For the weighted completion time scheduling problem, $\rho = 2$, $\gamma_1 = 3$ using cost budget $3C$, and $\gamma_2 = 1$. This implies an approximation ratio of $\eta = 10$ for the non-adaptive probing strategy that uses cost budget $3C$.
 \end{theorem}

The proof that $\gamma_2 = 1$ is simple, since the optimal scheduling
policy whose value is $\min_{o \in \O} \E_{\v_{\ol{S}}}[h(o; \v_S,
  \v_{\ol{S}})]$ will simply order the jobs in decreasing order of the
ratio of weight to (expected) processing time ({\em Smith's Rule}) --
this processing time is exactly known for probed jobs $S$, and is the
expected processing time for unprobed jobs $\ol{S}$. We will first
show $\rho = 2$, and then design a 3-approximation for the outlier problem that violates the cost budget by a factor of $3$.
 
\subsection{The Recombinant Property (P2)} 
We first show Property (P2) holds for $\rho = 2$.  Consider $n$ jobs with {\em deterministic} lengths. Let $l_1, l_2, \ldots, l_n$ denote the job-lengths, and $w_1, w_2, \ldots, w_n$ denote the job weights. Let $\alpha_i = \frac{l_i}{w_i}$. By Smith's rule, the optimal solution sorts the jobs in increasing order of $\alpha_j$ and schedules in this order. The completion time of the optimal ordering can therefore be written as $\displaystyle \sum_{i=1}^n \sum_{j \ge i} w_i w_j \min(\alpha_i, \alpha_j)$. We have:

\begin{lemma}
\label{lem:split}
For any partitioning of $n$ deterministic  jobs into two disjoint sets $A$ and $B$, we have:
$$  \sum_{i, j \in A, j \ge i} w_i w_j \min(\alpha_i, \alpha_j) +  \sum_{i, j \in
  B, j \ge i} w_i w_j \min(\alpha_i, \alpha_j) \ge  \sum_{i \in A, j \in B} w_i w_j \min(\alpha_i, \alpha_j)$$
\end{lemma} 
\begin{proof}
Let $\gamma_{ij} = \min(\alpha_i, \alpha_j)$. 
First, suppose all $\gamma_{ij} = 1$. We have:
\begin{eqnarray*}
& \left(\sum_{i \in A} w_i  - \sum_{j \in B} w_j\right)^2 & \ge \  0\\
\Rightarrow & 2\left(\sum_{i,j \in A, j \ge i} w_i w_j + \sum_{i,j \in B, j
\ge i} w_i w_j\right) & \ge \  2 \left(\sum_{i \in A} w_i\right)\left( \sum_{j \in B} w_j\right) \\
\Rightarrow & \sum_{i,j \in A, j \ge i} w_i w_j + \sum_{i,j \in B, j
  \ge i} w_i w_j & \ge \  \sum_{i \in A, j \in B} w_i w_j 
\end{eqnarray*}

We next consider the case of general $\alpha_i$. We will prove this by
induction on the number of jobs (the base case being trivial).  Let
$i^* = \mbox{argmin}_i \alpha_i$.  For each job $i$, let $\beta_i =
\alpha_i - \alpha_{i^*}$, and let $\delta_{ij} = \min(\beta_i,
\beta_j)$. We have:
$$ w_i w_j \gamma_{ij}  = w_i w_j  (\alpha_{i^*} +  \delta_{ij})$$
From the proof of the $\gamma_{ij} = 1$ case, we have:
$$\alpha_{i^*}\left(\sum_{i,j \in A, j \ge i} w_i w_j + \sum_{i,j \in B, j
  \ge i} w_i w_j\right) \ge \alpha_{i^*} \sum_{i \in A, j \in B} w_i w_j $$
The set of jobs with non-zero $\beta$ values is strictly smaller than $n$. Let  $Z$ be the set of jobs with $\beta_i =0$. By the inductive hypothesis we have:
$$\sum_{i, j \in A \setminus Z, j \ge i} w_i w_j \delta_{ij} +
\sum_{i, j \in B \setminus Z, j \ge i} w_i w_j \delta_{ij} \ge  \sum_{i \in A \setminus Z, j \in B \setminus Z} w_i w_j \delta_{ij}$$
Adding  the previous two inequalities, we have the proof of the lemma.
\end{proof}

In the above lemma, the LHS represents the contribution to the optimal completion time which arises from job pairs {\em within} $A$ and within $B$. The RHS represents contributions of job pairs such that one of the jobs is in $A$ and the other in $B$.  The above shows that the interaction term across the two sides of any partition, can be bounded by the sum of the interactions within each side.  This directly shows the recombinant property (P2) with $\rho = 2$:

\begin{lemma} 
$$ \forall A \subseteq [n], \v_A, \qquad  \min_{o \in \O} \E[h(o; \v_A, \v_{\ol{A}})] \le 2 \left[ \min_{o \in \O_A} h_A(o; \v_A) + \min_{o \in \O_{\ol{A}}} \E[h_{\ol{A}}(o; \v_{\ol{A}}) \right]$$
where all expectations are over $\v_{\ol{A}}$. 
\end{lemma}
\begin{proof}
Let $\mu_i = \E[X_i]$. We have: 
$$\min_{o \in \O_A} h_A(o; \v_A)  =  \sum_{i ,j \in A, j \ge i} \min(w_j \v_i, w_i \v_j)$$ 
$$\min_{o \in \O_{\ol{A}}} \E[h_{\ol{A}}(o; \v_{\ol{A}})  =  \sum_{i ,j \in \ol{A}, j \ge i} \min(w_j \mu_i, w_i \mu_j)$$ 
Furthermore, we have:
\begin{eqnarray*}
&&\min_{o \in \O} \E[h(o; \v_A, \v_{\ol{A}})]  \\
&& = \sum_{i ,j \in A, j \ge i} \min(w_j \v_i, w_i \v_j) +  \sum_{i ,j \in \ol{A}, j \ge i} \min(w_j \mu_i, w_i \mu_j) + \sum_{i \in A, j \in \ol{A}} \min(w_j \v_i, w_i \mu_j) \\
&&= \min_{o \in \O_A} h_A(o; \v_A) + \min_{o \in \O_{\ol{A}}} \E[h_{\ol{A}}(o; \v_{\ol{A}})  +  \sum_{i \in A, j \in \ol{A}} \min(w_j \v_i, w_i \mu_j)
\end{eqnarray*}
By Lemma~\ref{lem:split}, we have:
$$  \sum_{i \in A, j \in \ol{A}} \min(w_j \v_i, w_i \mu_j) \le   \sum_{i ,j \in A, j \ge i} \min(w_j \v_i, w_i \v_j) +  \sum_{i ,j \in \ol{A}, j \ge i} \min(w_j \mu_i, w_i \mu_j)$$
Putting the above two inequalities together completes the proof.
\end{proof}

\newcommand{\p}{\mathcal{P}}
\newcommand{\w}{\mathcal{W}}
\newcommand{\M}{\mathcal{M}}

\subsection{Approximation Algorithm for the Outlier Problem}
To complete the proof of Theorem~\ref{thm1}, we finally focus on
solving the outlier problem $\G^*(C)$.  Consider the following integer program for the outlier program. Here, variable $z_i \in \{0,1\}$ is set to $1$ if job $i$ is in the outlier set. Variable $e_{ij} \in \{0,1\}$ is set to $1$ if jobs $i$ and $j$ are both not in the outlier set.  
$$ \min \ \  \sum_{i=1}^n \sum_{j > i} e_{ij}\min\{w_j \mu_i, w_i \mu_j\} +  \sum_{i=1}^n (1-z_i) w_i \mu_i$$
\begin{eqnarray*}
& & \sum_{i=1}^n z_i c_i \leq C  \\
& & e_{ij} + z_i + z_j \geq 1  \mbox{~~for all $i,j > i$} \\ 
& & e_{ij}, z_i  \in \{0,1\}  \mbox{~~for all $i,j \in \{1,2,\ldots,n\}$} 
\end{eqnarray*}

It is easy to check that this integer program encodes the problem $\G^*(C)$. To see this, note that the objective is precisely the weighted completion time of the jobs not in the outlier set. The first constraint encodes the cost constraint on the outlier set, and the second constraint encodes that if jobs $i$ and $j$ are both not in the outlier set, then $e_{ij} = 1$, but if $z_i$ or $z_j$ is $1$, then $e_{ij} = 0$.

We solve the linear relaxation of the LP, where the final constraint is replaced with $e(i,j) \ge 0$ and $z_i \ge 0$.  Now round the LP solution as follows: Let $S =\{ i | z_i \geq \frac13\}$. Set $\tilde{z}_i = 1$ if $i \in S$, and set $\tilde{e}_{ij}=1$ if $i \not  \in S$ and $j \not \in S$. The set $S$ is the outlier set output by the algorithm.

\begin{theorem}
The outlier problem $\G^*(C)$ admits to a $3$-approximation that violates the cost budget $C$ by a factor of $3$.
\end{theorem}
\begin{proof}
It is easy to check that $\tilde{z}_i \le 3 z_i$ and $\tilde{e}_{ij} \le 3 e_{ij}$. To see this, simply observe that $e_{ij} = 1$ iff $z_i < 1/3$ and $z_j < 1/3$, in which case $e_{ij} \ge 1/3$. Therefore, the objective is within a factor $3$ of the LP objective, and the first constraint is violated by a factor of $3$.
\end{proof}

\section{Scheduling II: {\sc Makespan} on Identical Machines} 
\label{app:makespan}
We now consider the problem of minimizing the makespan on identical parallel machines. In this problem, there are $m$ identical machines, and $n$ jobs, whose sizes are random variables, $X_1, X_2, \ldots, X_n$. We can probe job $i$ by spending cost $c_i$, and find the exact value of its processing time.  Given a bound on the total query cost $C$, the goal is to find the subset of variables to probe so that the expected value of the makespan ({\em i.e.} the load on the most loaded machine) is minimized. Here, the expectation is over all realizations of the sizes of the unprobed jobs.

Recall that $\rho$ is the approximation for Property (P2), $\gamma_1$ is the approximation ratio for the outlier problem $\G(C^*)$, and $\gamma_2$ is the approximation ratio for solving $\min_{o \in \O} \E_{\v_{\ol{S}}}\left[h(o; \v_S, \v_{\ol{S}})\right]$. We will show the following theorem.

\begin{theorem}
\label{thm2}
For the minimum makespan scheduling problem, we have $\gamma_1 = O(1)$, $\gamma_2 = O(1)$, and $\rho = 1$, so that the approximation ratio of the non-adaptive strategy is $\eta = O(1)$. 
\end{theorem}

\paragraph{Recombinant Property.} It is easy to see that for any $A \in [n]$, for all $o_1 \in \O_A$ and $o_2 \in \O_{\ol{A}}$, the concatenation of these solutions (call it $o \in \O$) satisfies: 
$$ h(o; \v) \le h(o_1; \v_A) + h(o_2; \v_{\ol{A}}) \qquad \forall \v$$
This directly implies Property (P2) with $\rho = 1$.

\renewcommand{\P}{\mathcal{P}}

\paragraph{Outlier Problem.} The outlier version $\G^*(C)$ is the following: Find a subset $S$ of jobs with cost at most $C$ to discard so that the optimal expected makespan on the remaining jobs is minimized. We therefore consider the problem $\P$ of scheduling the jobs in the absence of probing. The goal is simply to assign the jobs to the machines so that the expected makespan, which is the expected load of the most loaded machine, is minimized. Here, the expectation is over all possible realizations of the job sizes. This problem was addressed by Kleinberg, Rabani, and Tardos~\cite{KleinbergRT97}, who show a nice characterization. 

\begin{define}
Given a random variable $X_i$, define the random variable
$Y_{i,t},Z_{i,t}$ as follows
\[ Y_{i,t} = \left\{ \begin{array}{ll} X_i/t & \mbox{if $X_i \leq t$}\\
0 & \mbox{otherwise} \end{array} \right. \quad \quad
Z_{i,t} = \left\{ \begin{array}{ll} 0 & \mbox{if $X_i \leq t$}\\
X_i & \mbox{otherwise} \end{array} \right. \]
In other words $Y$ is the contribution of $X$ below a threshold $t$ and $Z$ 
is the contribution of $X$ above $t$; naturally $X_i = t \cdot Y_{i,t} +
Z_{i,t}$.
Let 
\[ f_i(t) = \sum_i \left( E[Z_{i,t}] + \frac{t}{m} \frac{\log
    E[m^{Y_{i,t}}]}{\log m} \right) \]
\end{define}

\begin{theorem}[\cite{KleinbergRT97}]
\label{thm:krt}
Given a set of jobs whose sizes follow independent random variables $\{X_i\}$, 
\begin{itemize}
\item If $\sum_i f_i(t) \geq t/3$ then any scheduling of the jobs on $m$ machines
has expected makespan $\Omega(t)$.  
\item If $\sum_i f_i(t) \leq 2t/3$ then the following algorithm produces a schedule with (expected) makespan $O(t)$:
\begin{enumerate}
\item Define the ``effective size'' of job $i$ to be $\eta_i(t) = \frac{\log E[m^{Y_{i,t}}]}{\log  m}$.
\item Consider the jobs in arbitrary order, and place each job on that machine where the sum of the $\eta_j(t)$  of the already scheduled jobs $j$ is minimum. In other words, use Graham's rule on $\eta_j(t)$.
\end{enumerate}
\end{itemize}
\end{theorem}

The following algorithm solves $\G^*(C)$ to a $O(1)$. Try different values of $t$ in increasing powers of $(1+\EE)$. For each $t$, decide if there is a subset $S$ of jobs so that: (1) $\sum_{i \in S} c_i \le C$, and (2) $\sum_{i \notin S} f_i(t) \le t/2$. By Theorem~\ref{thm:krt}, the smallest such $t$ (call it $t^*$) for which the decision problem returns a ``yes" answer is the desired $O(1)$ approximation to $\G^*(C)$.  It is easy to see that $t^*$ will be at most the sum of the maximum values the $X_i$ can take.  Therefore, since the distributions are discrete and specified as input, computing $f_i(t)$ takes polynomial time for any $t$ of interest.  This decision problem is a knapsack problem, and can be solved to a $(1+\EE)$ approximation on $t^*$ without violating the cost budget. This shows $\gamma_1 = O(1)$. 

\medskip To show that $\gamma_2 = O(1)$, we simply  schedule the probed jobs using Graham's rule on the observed sizes. Separately schedule the unprobed jobs using the algorithm from~\cite{KleinbergRT97}, and append this schedule to that of the probed jobs. Since the probed jobs are scheduled to a factor $2$ approximation, and the unprobed jobs are scheduled to a $O(1)$ approximation (by the result of~\cite{KleinbergRT97}), this shows that $\gamma_2 = O(1)$, completing the proof of Theorem~\ref{thm2}.

\renewcommand{\P}{\mathcal{P}}
\newcommand{\V}{\mathcal{V}}
\renewcommand{\M}{\mathcal{M}}

\section{Metric Problems I: $K$-median Clustering}
\label{kmedian}
We next consider several metric problems, specifically the $k$-median clustering and the minimum spanning tree problem on metric spaces. Before defining the problems, we will define the uncertainty model.

\subsection{Uncertainty Model for Metric Problems}
\label{sec:uncle}
We define the uncertainty model as follows.  We are given a metric space with point set $\P$, which defines a distance function $l$. The input is a set of nodes $\V$, where the location of node $i$ follows an independent distribution $X_i$ over $\P$. Distribution (or node) $i \in \V$ has probe cost $c_i$; on probing, the node resolves to one of the locations in $\P$. Since we can only probe a subset of nodes, but must finally construct a solution over all the nodes, it will be helpful to define a metric space over the set of points $\P$ and nodes $\V$, {\em i.e.}, over the set $\M = \P \cup \V$. For $i,j \in \V$, let $D(i,j) = l(X_i,X_j)$ denote the random variable corresponding to the distance between $X_i$ and $X_j$. Let $d(i,j) = \E[D(i,j)]$ where the expectation is over the random variables $X_i$ and $X_j$.

\begin{claim}
\label{clmet}
$d(i,j) + d(j,k) \ge d(i,k)$.
\end{claim}
\begin{proof}
For any realization of the values of $X_i, X_j$, and $X_k$, we have $D(i,j) + D(j,k) \ge D(i,k)$. Taking expectations over the random choices of $X_i, X_j$, and $X_k$, the claim follows.
\end{proof}

Define the following metric space: The vertices are the points $\M = \P \cup \V$. The distance metric is $d$. For $i,j \in \V$, $d(i,j) = \E[D(i,j)] = \E[l(X_i,X_j)]$. For $i,j \in \P$, $d(i,j) = l(i,j)$. For $i \in \P, j \in \V$, define $d(i,j) = \E[l(i,X_j)]$. Using the above claim, it is clear that the function $d$ defines a metric space on the nodes $V \cup \P$.  This completes the modeling of uncertainty for the metric problems we consider.

\subsection{$K$-medians: Problem Statement}
In this problem, we are given a metric space with point set $\P$, which defines a distance function $l$. The input is a set of nodes $\V$, where the location of node $i$ follows an independent distribution $X_i$ over $\P$. Distribution $i \in \V$ has probe cost $c_i$. The goal is to design an adaptive policy to probe the nodes which spends expected cost at most $C$. After probing, the algorithm opens $K$ centers and assigns all probed and unprobed nodes to some center so that the expected distance cost (or {\em value}) of the clustering is minimized. This expectation is over  the  locations of the unprobed nodes. Note that after probing, the center selection and assignment policy assigns an unprobed node $i$ to that open center $w$ which minimizes the expected distance $\E[l(X_i,w)]$ where the expectation is  over the random variable $X_i$. The goal is to design a probing policy whose resulting expected distance cost (or value) of $K$-median clustering is minimized, where the expectation is over the outcomes of the probes and the locations of the unprobed nodes.

We consider two variants of the problem. In the first variant, we assume that the center selection policy is restricted  to opening centers from a set $S \subseteq \mathcal{P}$. This means the centers can only be chosen from points of the underlying metric space. Therefore, for an unprobed node $i$ assigned to a center $w \in \mathcal{P}$, the expected distance cost is $\E[l(X_i,w)]$, where the expectation is over the random variable $X_i$. In the second variant, the centers are allowed to be {\em input nodes}, and therefore distributions. Now, an unprobed node $j$ can be opened as a center after probing a set of nodes. Suppose an unprobed node $i$ is assigned to this center, then the expected distance cost is $\E[l(X_i,X_j)]$, where the expectation is over {\em both} the random variables $X_i$ and $X_j$. 

We present a constant factor adaptivity gap for the former variant.  We then show that the adaptivity gap is  polynomially large for the latter variant. This shows a fundamental difference in the two variants.

\subsection{Fixed Centers}  We first consider the first variant where the centers can only be points from $S \subseteq \mathcal{P}$.  Recall that $\rho$ is the approximation for Property (P2), $\gamma_1$ is the approximation ratio for the outlier problem $\G(C^*)$, and $\gamma_2$ is the approximation ratio for solving $\min_{o \in \O} \E_{\v_{\ol{S}}}\left[h(o; \v_S, \v_{\ol{S}})\right]$. We will show the following theorem.

\begin{theorem}
\label{thm3}
For the $K$-median problem with fixed centers, $\rho = 5$, $\gamma_1 = 5$ with outlier cost $5C$, and $\gamma_2 = 3+\EE$. This shows an approximation ratio of $\eta = O(1)$ with outlier cost $5C$.
\end{theorem}

As shown in Section~\ref{sec:uncle}, first define the new metric space $d$ over the points $\P \cup \V$.  Given the set of probed values $\v_S \subseteq \P$, and the unprobed nodes $\bar{S} \subseteq \V$,  the problem reduces to deterministic $K$-medians on $\v_{S} \cup \bar{S} \subseteq \P \cup \V$ in metric space $d$, and this shows $\gamma_2 = 3+\EE$ using the algorithm in~\cite{AryaGKMMP04}. In the outlier problem, the goal is to find the subset $T \subseteq \V$ of nodes of total probing cost at most $C$ such that the cost of $K$-median clustering of the remaining nodes $\V \setminus T$ in metric space $d$ is minimized. This problem has a $\gamma_1 = 5$ approximation due to the authors of~\cite{CharikarKMN01} if the outlier cost is relaxed to $5C$.  If all probing costs are unit, we have $\gamma_1 = O(1)$ approximation 
that does not relax the outlier cost \cite{kechen}. The only part left to show is the recombinant property (P2).

\newcommand{\F}{\mathcal{F}}

\begin{lemma}
$$\forall A \subseteq [n], \v_A, \ \ \min_{o \in \O} \E\left[h(o; \v_A, \v_{\ol{A}})\right]  \le 5 \cdot  \min_{o \in \O_{\ol{A}}} \E[h_{\ol{A}}(o; \v_{\ol{A}})] + 4 \cdot \E[\min_{o \in \O} h(o; \v_A, \v_{\ol{A}})] $$
where the expectations are over $\v_{\ol{A}}$.
\end{lemma}
\begin{proof}
The quantity $Q_1 =  \min_{o \in \O} \E\left[h(o; \v_A, \v_{\ol{A}})\right] $ on the LHS involves constructing the optimal $K$-median solution using the actual realization $\v_A$ of the probed set $A$, and the expected locations of the unprobed nodes $\ol{A}$. To bound this cost, we first construct the optimal $K$ medians on the unprobed nodes $\ol{A}$ by paying $Q_2 = \min_{o \in \O_{\ol{A}}} \E[h_{\ol{A}}(o; \v_{\ol{A}})]$.   Next,  move the unprobed nodes to the assigned median. For $i \in \ol{A}$, let $\phi(i)$ denote the assigned median.

We will now proceed by considering scenarios $\v_A, \v_{\ol{A}}$ of the values of all nodes. Fix some such scenario $\sigma$. Let $M_{\sigma}$ denote the value of the optimal $K$-median solution in this scenario. The expected value of this optimal solution over the realizations of $\v_{\ol{A}}$ is $Q_3 = \E[\min_{o \in \O} h(o; \v_A, \v_{\ol{A}})] $. Let $P' \subseteq \P$ denote the set of points $\v_A$, along with the nodes $i \in \ol{A}$ located at point $\phi(i)$. Note that the set $P_{\sigma}$ is the same for all realized $\sigma$. Since $\v_A$ is fixed, and the mapping $\phi$ is independent of the scenario, we have $P'$ is independent of the scenario $\sigma$.

We will now construct the solution corresponding to the quantity $Q_1$ using the solution for $Q_2$ and the solutions $M_{\sigma}$.  In scenario $\sigma$,  send each point in $A$ its assigned center in the optimal solution $M_{\sigma}$.  Similarly, send each node $i \in \ol{A}$ located at $\phi(i)$ back to its realized location in scenario $\sigma$ and from there to its assigned center in $M_{\sigma}$. This yields a mapping from the nodes $P'$ to $K$ medians in scenario $\sigma$. The expected distance value (over scenarios $\sigma$) of this mapping is at most $Q_2 + Q_3$ by triangle inequality. Note that the distances in this mapping are distances between points in $\P$, and do not involve the distributional nodes $\V$. This yields a valid $K$-median solution on $P'$.

Since there is a feasible $K$-median solution for each $\sigma$, these when averaged over $\sigma$ define a fractional $K$-median solution for the set of points $P'$ of value at most $Q_2 + Q_3$. Therefore, there is an integer $K$-median solution on $P'$ of value $4$ times this fractional value~\cite{CharikarG99}. The final solution corresponding to $Q_1$ maps the points $i \in \ol{A}$ first to $\phi(i)$ paying cost $Q_2$, and then maps all the points to these constructed centers paying cost $4(Q_2 + Q_3)$. This shows $Q_1 \le 5 Q_2 + 4 Q_3$, completing the proof.
\end{proof}

\subsection{Arbitrary Centers}
Consider now the case where the centers themselves are allowed to be input nodes, and therefore distributions. After probing, the center selection policy could decide to open an unprobed input node as a center, and assign probed and unprobed nodes to this center. The distance cost between the center and the assigned node is the expected distance between them, where the expectation is taken over possible locations of the center and the assigned node. We note that the linear programming relaxation is still a lower bound on the adaptive optimal solution.

By constructing separate $K$-median solutions for the probed and unprobed variables, it is easy to show a $O(1)$ approximation that uses $2K$ centers and pays probing cost $5C$. We show that this is the best possible result for this case in the following sense. We show that any non-adaptive algorithm has polynomially large adaptivity gap on {\em both} distance and probing cost if it is restricted to opening at most $(1+\epsilon)K$ centers for some constant $\epsilon > 0$.  Intuitively what fails in reducing the $2K$ centers to $K$ centers is the following: In the proof for fixed centers, we mapped the unprobed nodes of the non-adaptive solution to a set of nodes in $\P$, and constructed a fractional solution using these locations for these nodes. This ensures that the metric space is over $\P$ and independent of distributions. In the current setting, an unprobed node could be mapped to an unprobed center. Therefore across scenarios, the metric space over the locations of the nodes changes.

\begin{theorem}
The adaptivity gap for $K$-medians when centers can be input nodes is
polynomially large on both distance and probing cost.
\end{theorem}
\begin{proof}
 Consider $M$ distinct copies (at a mutual distance of at least $M^3L$
  from each other, $L=M^2$) of the following $2$-dimensional node
  set. In copy $m$, there are $r+1$ ``cheap'' nodes $X_{1m}, X_{2m},
  \ldots$ which cost $1$ to probe.  Distribution $X_{im}$ is $(0,0)$ with
  probability $1/2$ and $(i+1,0)$ otherwise. In addition, there are $t \gg \sqrt{M}$ pairs of
  nodes which are well-separated from other pairs by a large distance
  $L$.  Pair $j$ corresponds to two distributions: $Y_{jm}$ and
  $Z_{jm}$.  $Y_{jm}$ is $(L+jL,1)$ with probability $1-\log t/t$ and
  $(L+jL,0)$ with probability $\log t/t$. $Z_{jm}$ is $(L+jL,-1)$ with
  probability $1-\log t/t$ and $(L+jL,0)$ with probability $\log
  t/t$. These distributions are ``expensive'' with probing cost
  $(r+1)M$, where $r > 2 \log Mt$. Again note that the nodes for each $m$ are far removed
  from the nodes corresponding to other $m$.

Let $K=(2t+r)M$.  For each $m$, the adaptive solution will place $2t+r$
medians using the following strategy.  First probe all cheap distributions $X_{*}$. There are two cases:

\medskip
\noindent {\bf Case 1.} For some $m$, $X_{*m}$ resolve to $r+1$ values distinct points. This
happens with probability $1/2^{r+1}$ for a particular $m$ and
therefore $\frac{M}{2^{r+1}}$ overall.  In this case,   probe {\em all} the
``expensive'' distributions $Y_{*}$ and $Z_{*}$.  Choose the $r+1$ realized points $X_{*}$, and some $2t-1$ of the remaining $2t$ points as medians.  With probability
$1-(1-\log t/t)^{2t} \ge 1 - 1/t^2$ some $(Y,Z)$ pair collides on the
$x$-axis. In this case, the $k$ medians solution has distance cost
$0$, else it has a distance cost of at most $2$. The expected
distance cost is therefore at most $(1/2^{r+1})\cdot (1/t^2) \cdot 2$
for each $m$. Therefore, the overall expected distance cost is 
$\frac{M}{t^22^r}$. 

\medskip
\noindent {\bf Case 2.} If at most $r$ distinct values of $X_{*m}$ are observed for all $m$, then do not probe further,
since there is a $k$ median solution of value $0$, in which every one
of the expensive nodes and all the realized locations of the cheap nodes
are the medians. 

\medskip
The expected probing cost of this scheme is at most
$M(r+1) + (r+1)M \cdot 2Mt \cdot \frac{M}{2^{r+1}} \le 4M (r+1)$.

\medskip
Any non-adaptive probing scheme must probe at least one expensive
distribution in each copy, else the distance cost is at least
$\frac{1}{2^{r+2}} \gg \frac{M}{t^22^r}$ in that copy: In the case
where the cheap distributions resolve to distinct values
w.p. $1/2^{r+1}$, the distance cost will be at least
$0.5$. Therefore, the probing cost needed is $(r+1)M^2$, which
implies that unless the probing cost is a factor $M$ larger than the
adaptive scheme, the distance cost must be $\frac{t^2}{M}$ times
larger. Therefore, no non-trivial adaptivity gap is possible.
\end{proof}

\section{Metric Problems II: Minimum Steiner Trees}
\label{app:mst}
We finally consider the minimum Steiner tree problem -- the algorithm extends to Metric Traveling salesman problem naturally. As with $K$-medians, the input is a collection of $n$ nodes. The location of node $i$ is an independent random variable $X_i$, which is a distribution over points $\mathcal{P}$ in a metric space with distance function $l$. Let $\V$ denote the set of nodes. The exact location of node $i$ is determined by spending probing cost $c_i$. The goal is to design an adaptive probing scheme which minimizes the expected cost of connecting the nodes by a spanning tree, subject to the constraint that this decision tree has expected probing cost at most $C$.  

Recall that $\rho$ is the approximation for Property (P2), $\gamma_1$ is the approximation ratio for the outlier problem $\G(C^*)$, and $\gamma_2$ is the approximation ratio for solving $\min_{o \in \O} \E_{\v_{\ol{S}}}\left[h(o; \v_S, \v_{\ol{S}})\right]$. We will show the following theorem.

\begin{theorem}
For the metric Steiner tree problem, we have $\rho = 1$, $\gamma_1 = 4$ with outlier cost $4C$, and $\gamma_2 = 1.55$. The non-adaptive strategy therefore has an approximation ratio $\eta = O(1)$ with probing cost $4C$. 
\end{theorem}

Since the overall algorithm is similar to the $K$-medians problem, we simply highlight the differences. As discussed in Section~\ref{sec:uncle}, we construct the metric space $d$ over $\M = \P \cup \V$. The outlier problem  over the nodes $\V$ in the metric space $d$ is defined as follows: There is a cost of $c_i$ for every node $i \in \V$. The goal is to choose a set of nodes $S \subseteq \V$ with total cost at most $C$ to discard such that the value of the Steiner tree on the nodes in $\V \setminus S$ is minimized. This has a $\gamma_2 = 4$-approximation
that spends outlier cost $4C$ \cite{GoemansW}.  If all probing costs are unit, we have $\gamma_2 = 4$-approximation~\cite{garg} that obeys the cost constraint. The minimum cost Steiner tree problem on the probed and unprobed nodes has an approximation ratio of  $\gamma_2 = 1.55$~\cite{zelikovsky}. 

To show the recombinant property (P2) with $\rho = 1$, we make the following observations.  For any set $A$ of probed nodes with realization $\v_A$,  $h_A(o; v_A)$ is the value of the optimal Steiner tree $T_1$ connecting these realized nodes; $\min_{o \in \O_{\ol{A}}}\E[h_{\ol{A}}(o; \v_{\ol{A}})]$ is the value optimal Steiner tree $T_2$ connecting the unprobed nodes; and finally, $\E[\min_{o \in \O} h(o; \v_A, \v_{\ol{A}})]$ is at least the expected cost of any edge $e =(i,j)$ connecting some $i \in A$ and $j \in \ol{A}$. Taking the union of $T_1$, $T_2$, and $e$  yields a tree whose cost is at least $\min_{o \in \O} \E[h(o; \v_A, \v_{\ol{A}})]$. 

\section{Conclusion}
In this paper, we considered a large class of scheduling and metric problems when the inputs follow distributions that can be observed by paying a cost. We showed that the problem of computing the optimal budgeted resolution scheme for these inputs (that minimizes the expected value of the posterior optimization) is closely related to the outlier version of these problems. This work results in several open questions. First, can the approximation ratios be improved by an algorithm that designs adaptive resolution schemes instead of non-adaptive schemes? Next, our model for metric problems assumes nodes are uncertain and can be resoled at a cost. What if only {\em distances} between nodes can be resolved at a cost? The problem now is that the distributions for the edge lengths cannot be independent random variables, and we would need new techniques. Finally, it would be interesting to explore if the outlier scheme can be extended to a larger class of model-driven optimization problems. 

\medskip
\noindent {\bf Acknowledgments:} We thank the anonymous referee for significantly improving the presentation of this paper. We also thank Sariel
Har-Peled, Ashish Goel and Anupam Gupta for helpful discussions.

\bibliographystyle{plain}

\end{document}